\documentclass[preprint,prl,superscriptaddress,amsmath,amssymb]{revtex4-2}
\usepackage{amsmath}
\usepackage{amsfonts}
\usepackage{amssymb}
\usepackage{graphicx}
\usepackage{color}
\usepackage{txfonts}
\usepackage[titletoc]{appendix}
\usepackage[colorlinks={true}]{hyperref}
\hypersetup{citecolor={blue}, filecolor={blue}, linkcolor={blue}, urlcolor={blue}}
\begin{document}
\preprint{}
%\title{Cavity-modified field-free orientation of a single molecule}
%\title{Terahertz Pulse-Induced Orientation of a Single Molecular Polariton}
\title{Quantum Coherent Control of a Single Molecular-Polariton Rotation}
\author{Li-Bao Fan}
\affiliation{Hunan Key Laboratory of Nanophotonics and Devices, Hunan Key Laboratory of Super-Microstructure and Ultrafast Process, School of Physics and Electronics, Central South University,
Changsha 410083, China}
\author{Chuan-Cun Shu}
\email{cc.shu@csu.edu.cn}
\affiliation{Hunan Key Laboratory of Nanophotonics and Devices, Hunan Key Laboratory of Super-Microstructure and Ultrafast Process, School of Physics and Electronics, Central South University,
Changsha 410083, China}
\author{Daoyi Dong}
\affiliation{School of Engineering and Information Technology, University of New South Wales,
Canberra, Australian Capital Territory 2600, Australia}
\author{Jun He}
\affiliation{Hunan Key Laboratory of Nanophotonics and Devices, Hunan Key Laboratory of Super-Microstructure and Ultrafast Process, School of Physics and Electronics, Central South University,
Changsha 410083, China}
\author{Niels E. Henriksen}
\affiliation{Department of Chemistry, Technical University of Denmark, Building 207, DK-2800 Kongens Lyngby, Denmark}
\author{Franco Nori}
\affiliation{Theoretical Quantum Physics Laboratory, RIKEN, Saitama 351-0198, Japan}
\affiliation{Physics Department, University of Michigan, Ann Arbor, Michigan 48109, USA}
 \begin{abstract}
 We present a combined analytical and numerical study for coherent terahertz control of a single molecular polariton, formed by strongly coupling two rotational states of a molecule with a single-mode cavity. Compared to the bare molecules driven by a single terahertz pulse, the presence of a cavity strongly modifies the post-pulse orientation of the polariton,  making it difficult to obtain its maximal degree of orientation. To solve this challenging problem toward achieving complete quantum coherent control, we derive an analytical solution of a pulse-driven quantum Jaynes–Cummings model by expanding the wave function into entangled states and constructing an effective Hamiltonian. We utilize it to design a composite terahertz pulse and obtain the maximum degree of orientation of the polariton by exploiting photon blockade effects. This work offers  a new strategy to study rotational dynamics in the strong-coupling regime and provides a method for complete quantum coherent control of a single molecular polariton. It, therefore, has direct applications in polariton chemistry and molecular polaritonics for exploring novel quantum optical phenomena. 

\end{abstract}
\maketitle
Since the pioneering experimental work by Ebbesen and collaborators in 2012 \cite{ACIE:51:1592}, strong light-matter interactions between an internal mode of molecules and a cavity-photon mode have attracted substantial attention in chemistry and physics \cite{PRL2015,JPCL2016NaI,PNAS2017,JPCL2018Na,JPCL2018,JPCL2020,PRB20212}. It promptly established an emerging field within chemistry called polariton chemistry \cite{cs2018,LPR2019,csr2019,PRX2020,PNAS-2021,science-Ebbesen}, bringing together quantum optics and chemical physics.  Considerable experimental and theoretical works have demonstrated that strong light-matter coupling can significantly modify the physical and chemical properties of molecules \cite{PRX2015, PRL2019,JCP2020,PRB2021,PRL20212,JCP20212}.  Experimentally, it has become possible to achieve vibrational-strong coupling (VSC) of a single molecule with nanocavities  even at room temperature \cite{nature2016,nc2019}. The formation of vibrational polaritons as the  quantum superpositions of molecular vibrations and cavity-photon modes offers an alternative approach for selective activation of chemical bonds and an efficient pathway for energy transfer between molecules \cite{science2019,pnas2020,science2020,nc2021,JACS2021,JCP2021}.\\ \indent
The rotational degree of freedom of molecules can provide physical insights into light-matter interactions beyond the VSC limit. This work explores the rotational-strong coupling (RSC) and focuses on quantum control of molecular rotation \cite{qr-TS,IRPC2010,IJC2012,qr-sugny}. Recent experiments have shown how to orient single polar molecules on a surface with maximum precision using the electric field of a scanning turning microscope \cite{nc20192} and how to control the orientation of  molecules inside a cavity  for enhancing the VSC \cite{CEJ2017}. However, the generation of post-pulse orientation of a single molecule inside a cavity remains unexplored, which is related to a phenomenon known as field-free orientation \cite{qr-niels2}. In the absence of a cavity, there has been significant progress for generating post-pulse orientation of molecules using resonant terahertz pulses \cite{qr-niels2,sugny2,pra2006,shu3,shuJCP,ex0,ex00,jiro,shu4,ex1,ex2,PRL2020,shu2020,shupra2021}  or nonresonant optical pulses \cite{qr-kling,PRL:2014,qr-wj}. It is crucial to show how cavity quantum electrodynamics (c-QED) effects modify such a phenomenon as a direct consequence of RSC. \\ \indent
In this Letter, we  theoretically explore the realization of the maximal post-pulse orientation of a single polariton irradiated by pulsed terahertz fields, which is of much current interest to the existing techniques on molecular orientation and the field of single-molecule strong-coupling in cavities. We examine an archetypal model of RSC for two rotational states of a single molecule inside a single-mode cavity. Previously, by combining a long nonresonant laser pulse and a weak static electric field to mix the lowest two rotational states with equal weights, the maximal two-state-post-pulse orientation has been demonstrated experimentally for ultracold carbonyl sulfide (OCS) molecules without a cavity \cite{ts}. Here, we show that this orientation maximum is reachable for the bare molecule using a single terahertz pulse without using a static electric field. Furhermore, the orientation features are strongly modified when placing the molecule into a cavity, showing a direct signature of hybrid entangled states. By expanding the time-dependent wave function into entangled states and using an effective Hamiltonian, we derive an analytical solution of a pulse-driven quantum Jaynes-Cummings (JC) model  \cite{JCM1963}  to describe the molecular-polariton dynamics without the rotating wave approximation (RWA). We then utilize it to design a complete quantum coherent control scheme, which can achieve the orientation maximum of the polariton by controlling the amplitude and phase of a composite terahertz pulse while exploiting quantum blockade effects.\\ \indent
\begin{figure}[h]
\resizebox{0.5\textwidth}{!}{%
\includegraphics{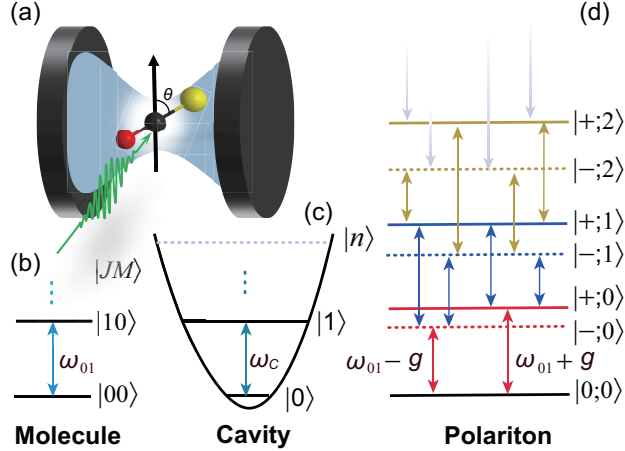} }\caption{(a) Schematic of a pulse-driven molecular-polariton and the corresponding  energy levels.  A single molecule with rotational states $|JM\rangle$ at a rotational frequency $\omega_{01}$  in (b) is strongly coupled with a single-mode cavity with photon states $|n\rangle$ at a frequency $\omega_c $ in (c), resulting in entangled states $|0;0\rangle$ and $|\pm;n\rangle$ in (d). $\theta$ denotes the angle between the molecular axis and the driving-field's polarization vector.}
\label{fig1}
\end{figure}

We consider a single polar molecule  initially in the absolute-ground state strongly coupled with a single-mode cavity of a frequency $\omega_c$ and  driven by a linearly polarized terahertz pulse  $\mathcal{E}(t)$, as illustrated in Fig. \ref{fig1}(a). By setting the the laser field's polarization vector $\mathbf{\hat{e}}$ along the cavity one, the total Hamiltonian of the molecule in the presence of the cavity and pulsed fields can be described in the dipole gauge  by  ($\hbar=1$) \cite{Yoshihara2016,Stefano2019,Taylor2020}
\begin{equation} \label{Hcm}
\hat{H}_{\mathrm{tot}}(t)=B\hat{J}^2+\omega_c\hat{a}^\dag\hat{a}-\sqrt{\frac{\omega_c}{2\epsilon_0 V}}\mathbf{\hat{\mu}}\cdot\mathbf{\hat{e}}\left(\hat{a}+\hat{a}^\dag\right) -\mathbf{\hat{\mu}}\cdot\mathbf{\hat{e}}\mathcal{E}(t),
\end{equation}
where the first term is the field-free rigid-rotor  Hamiltonian with
the angular momentum operator $\hat{J}$ and rotational constant $B$,  and the second one denotes the cavity field Hamiltonian with $\hat{a}^\dag$ ($\hat{a}$) being the creation (annihilation) photon operators. The last two terms in Eq. (\ref{Hcm}) describe the interactions of the molecule with a cavity and a laser field, respectively, where $\epsilon_0$, $V$, and $\mathbf{\hat{\mu}}$ represent the electric constant, volume of the electromagnetic mode, and  dipole operator of the molecule.  %Note that the model excludes the vibrational degree of freedom of the molecules due to its energy far above the frequencies of both the cavity model and terahertz pulses.
\\ \indent
In the absence of the second and third terms in Eq. (\ref{Hcm}), post-pulse orientation of the bare molecule can be generated by applying  resonant terahertz laser pulses \cite{shu2020,shupra2021}. By designing a single terahertz pulse  to satisfy the condition  $\left|\mu_{01}\int_{t_0}^{t_f}\mathcal{E}(t)\text{exp}[i\omega_{01}t]dt\right|=|\theta_{01}(t_f)|=\pi/4$ with  $\mu_{01}=\langle 00|\mathbf{\hat{\mu}}\cdot\mathbf{\hat{e}}|10\rangle$ and $\omega_{01}=2B$, while keeping its bandwidth narrow enough, we obtain a coherent superposition of rotational states  $|00\rangle$ and  $|10\rangle$ with equal weights without populating higher rotational states. The corresponding post-pulse orientation  reaches the  maximum of $\langle\cos\theta\rangle_{\mathrm{max}}=\sqrt{3}/3\approx0.5774$ with  the revival period of $\tau=\pi/B$, which can be modified by applying a  weak static field \cite{SM}. In the presence of strong molecule-cavity coupling, the third term in Eq. (\ref{Hcm}) mixes the bare molecular and photon states into hybrid entangled states in Figs. \ref{fig1}(b-d).  \\ \indent
\begin{figure}[h]
\resizebox{0.45\textwidth}{!}{%
\includegraphics{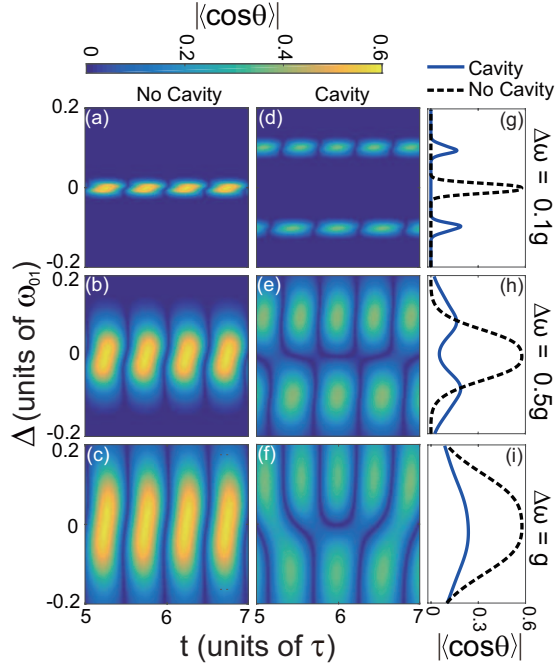} }\caption{The dependence of the numerically calculated time-dependent degree of orientation $|\langle\cos\theta\rangle(t)|$  on the laser frequency detuning $\Delta$ with time in units of the full revival period $\tau$ of the bare molecule. The  comparisons without (a)-(c) and with (d)-(f) a cavity are calculated with three different pulse-bandwidths $\Delta\omega=0.1g$ (top), $0.5g$ (middle), and $1.0g$ (bottom) while keeping other parameters unchanged.  (g)-(i)  The simulations of $\langle\cos\theta\rangle(t)$ versus $\Delta$  at a given time $t=6.75\tau$.}
\label{fig2}
\end{figure}
To examine how the c-QED affects the post-pulse orientation of the bare molecule, we consider OCS molecules at ultralow temperatures as an example with $B=0.20286$ cm$^{-1}$ and $\mu=0.715$ D. Figure \ref{fig2} shows the comparisons of the time-dependent degree of orientation $\langle\cos\theta\rangle(t)$  without and with a cavity. We consider the cavity mode in resonance with the transition between $|00\rangle$ and $|10\rangle$ with $\omega_c=\omega_{01}$ and set the strength of the polariton  interaction $g=\sqrt{\omega_c/(2\epsilon_0 V)}\langle 00|\mu\cos\theta|10\rangle=0.1\omega_{01}$. A Gaussian-profile pulse  $\mathcal{E}(t)=\mathcal{E}_0f(t)\cos(\omega_0t+\phi_0)$ is used, where $f(t)=\exp[-t^2/2\tau_0^2]$ and $\mathcal{E}_0=\sqrt{2/\pi}A_0/(\mu_{01}\tau_0)$ with  the duration $\tau_0$, center frequency $\omega_0$, and absolute phase $\phi_0$. This choice leads to $|\theta_{01}(t_f)|=A_0$ at the frequency $\omega_0$, independent of duration $\tau_0$. We examine the differences with the use of the pulse at three different  bandwidths $\Delta\omega=1/\tau_0=0.1g$, $0.5g$, and $1.0g$ while fixing $A_0=\pi/4$ and $\phi_0=0$. At each given $\Delta\omega$, we perform the simulations by scanning the frequency detuning $\Delta=\omega_0-\omega_{01}$, from $-0.2\omega_{01}$ to $0.2\omega_{01}$.\\ \indent
Figure \ref{fig2} shows that the polariton's post-pulse orientation phenomena are very distinct from the bare molecule one.  For the case $\omega_0=\omega_{01}$, the  orientation maximum 0.5774 appears in the bare molecule for different pulse bandwidths.  For the polariton, however, the post-pulse orientation is affected and even disappears using a narrow-bandwidth pulse. As $|\Delta|$ becomes large, the orientation maximum of the bare molecule decreases, whereas the polariton one becomes visible. Interestingly, the oscillation period of the polariton orientation becomes longer (shorter) for the negative (positive) detuning compared to the bare molecule. It implies that the manipulation of the molecule-driving pulse can selectively control the rotation speed of the molecule in a cavity.\\ \indent
\begin{figure}[h]
\resizebox{0.45\textwidth}{!}{%
\includegraphics{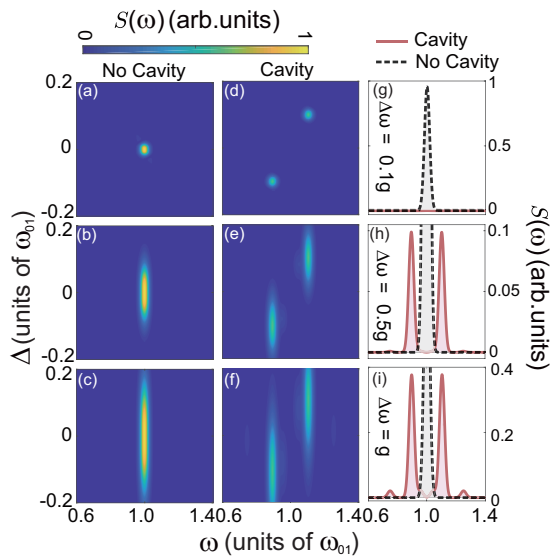} }\caption{The corresponding Fourier transform spectra $S(\omega)$ of $\langle\cos\theta\rangle(t)$ in Fig. \ref{fig2}. The  comparisons without   (a)-(c) and with (d)-(f) a cavity are calculated at three different laser bandwidths $\Delta\omega=0.1g$ (top), $0.5g$ (middle)  and $1.0g$ (bottom).  (g)-(i) The simulations of  $S(\omega)$ at detuning $\Delta=0$.}
\label{fig3}
\end{figure}
To understand the c-QED effect in Fig. \ref{fig2}, we perform an analysis using the hybrid entangled states \cite{JCM1963,JCM1993}. We treat the first three terms in Eq. (\ref{Hcm}) under the JC model Hamiltonian $\hat{H}_{\mathrm{JC}}=\omega_{01}|10\rangle\langle10|+\omega_c\hat{a}^\dag\hat{a}+g(|00\rangle\langle10|\hat{a}^\dag+\hat{a}|10\rangle\langle00|)$. By diagonalizing the  Hamiltonian $\hat{H}_{\mathrm{JC}}$ at the resonant cavity-molecule coupling $\omega_c=\omega_{01}$, we  obtain the polariton's eigenenergies $\omega_{0,0}=0$ and $\omega_{\pm,n}=\omega_c(n+1)\pm g\sqrt{n+1}$, associated with the vacuum ground state  $|0;0\rangle=|00\rangle|0\rangle$ and maximally entangled  states $|\pm;n\rangle=\sqrt{2}/2\left(|00\rangle|n+1\rangle\pm|10\rangle|n\rangle\right)$. The pulsed-driven polariton's Hamiltonian in  terms of hybrid entangled states reads,
\begin{eqnarray} \label{Hp}
\hat{H}_{\mathrm{p}}(t)&=&\sum_{n=0}^{\infty}\sum_{\ell=\pm}\omega_{\ell,n}|\ell;n\rangle\langle \ell;n|-\mathcal{E}(t)\sum_{\ell=\pm}\tilde{\mu}_{0}\Big(|\ell;0\rangle\langle0;0|+|0;0\rangle\langle\ell;0|\Big) \nonumber \\
&&-\mathcal{E}(t)\sum_{n=1}^{\infty}\sum_{\ell,\ell'=\pm}\tilde{\mu}_{\ell}\Big(|\ell;n\rangle\langle\ell';n-1|+|\ell';n-1\rangle\langle\ell;n|\Big),
\end{eqnarray}
where $\tilde{\mu}_0=\pm\sqrt{2}/2\mu_{01}$ and  $\tilde{\mu}_{\pm}=\pm1/2\mu_{01}$ denote the transition dipole moments between entangled states with $\mu_{01}=\langle 00|\mu\cos\theta|10\rangle=\sqrt{3}/3\mu$. The RSC splits the excited states $|\pm;n\rangle$ by the Rabi splitting $2\sqrt{n+1}g$ and modifies the dipole moments, as shown in Figs. \ref{fig1}(b)-(d). \\ \indent
Figure \ref{fig3} shows the Fourier-transform spectra of the calculated time-dependent degree of orientation in Fig. \ref{fig2}, i.e.,  $S(\omega)=\left|\int_{t_0}^{\infty}dt\langle\cos\theta\rangle(t)\exp(i\omega t)\right|$, which can be used to retrieve the transitions between the involved dressed states. For a resonant molecule-driving $\omega_0=\omega_{01}$ with  a narrow bandwidth  $\Delta\omega=0.1g$ in Figs. \ref{fig3}(a,d,g), the vacuum Rabi splitting blocks  one-photon transition from $|0;0\rangle$ to $|\pm,0\rangle$, leading to the disappearance of the orientation in Fig. \ref{fig2}(d) for $\Delta = 0$. By tuning $\omega_0$ away from $\omega_{01}$,  the two states $|\pm;0\rangle$ can be excited independently, resulting in an orientation enhancement at $\Delta=g$ and $-g$ in Fig. \ref{fig2}.  For the molecule-driving with  broad bandwidths of $\Delta\omega=0.5g$ and $1.0g$, the spectra in Figs. \ref{fig3}(e,h,f,i) consist of symmetric doublet peaks around the bare-molecule resonances. The first doublet peaks at $\omega_{01}\pm g$ in Figs. \ref{fig3}(e,h) implies that the laser field  described by the second term in Eq. (\ref{Hp}) drives the transitions from $|0;0\rangle $ to  $|\pm;0\rangle$ in Fig. \ref{fig1}(d) simultaneously. The higher-lying doublet peaks become visible in Figs. \ref{fig3}(f,i), corresponding to  the further transitions from $|\pm; 0\rangle$ to $|\pm;1\rangle$. The corresponding orientation phenomena  become much more complex in Figs. \ref{fig2}(e,f) than the  bare two-state  molecule in Figs. \ref{fig2}(b,c). It opens an important question of whether there exists a theoretical orientation maximum and how to achieve it for such a molecular rotational polariton.\\ \indent
To address this challenging question, we expand the time-dependent wave function $|\psi(t)\rangle$ into the hybrid entangled states
\begin{equation} \label{WFp}
|\psi(t)\rangle=\sum_{\ell=0,\pm}C_{\ell,0}e^{-i\omega_{\ell,0}t}|\ell;0\rangle+\sum_{n'=1}^{\infty}\sum_{\ell'=\pm}C_{\ell',n'}e^{-i\omega_{\ell',n'}t}|\ell';n'\rangle,
\end{equation}
 where $C_{\ell,0}$ denote the complex coefficients of the lowest three states $|0;0\rangle$ and $|\pm;0\rangle$,  and $C_{\pm,n}$  correspond to the complex coefficients of the higher-lying doublet states $|\pm;n\rangle$ with $n>0$. Based on the method of Lagrange multipliers \cite{Sugny:PRA:2004,NEH2020}, we conclude that the orientation maximum of the polariton is $\langle\cos\theta\rangle_{\mathrm{max}}=\sqrt{3}/3$, the same as for the bare molecule \cite{SM}. In this work, we achieve this maximum by  generating a coherent superposition  of the lowest three states with $|C_{0,0}|^2=0.5$, $|C_{+,0}|^2=|C_{-,0}|^2=0.25$, while satisfying a phase relation of  $\omega_{-,0}\arg[C_{+,0}]-\omega_{+,0}\arg[C_{-,0}]=2g\pi$. This implies that  $|C_{\pm,n'}|=0$ in the second term in  Eq. (\ref{WFp})  by  blocking the  transitions driven by the third term in Eq. (\ref{Hp}).\\ \indent
 Since the usual pulse area theorem with the RWA \cite{Ciappina2015,Fischer2018} fails to  include the pulse-phase information explicitly,  we derive an analytical solution of the time-dependent wave function within the pulse-driven quantum JC model described by Eq. (\ref{Hp}) without the RWA,  which remains challenging in quantum optics \cite{Schuster2008,Deppe2008,Liu2005,Fink2008,Bishop2008,Hamsen2017,PR2017}.  We consider the model including the lowest three states $|0;0\rangle$ and $|\pm;0\rangle$ and the  second higher-lying doublet states $|\pm;1\rangle$. We describe the transitions between the five states using an
effective three-level model with sublevels \cite{SM} and involving  the first-order Magnus expansion of the unitary operator \cite{shuprl2019,shuol2020,pra:92:063815,shu2020,shupra2021}.  The corresponding  time-dependent wave function  for the polariton starting from $|0;0\rangle$ reads \cite{SM}
\begin{eqnarray}\label{WFp1}
\vert\psi^{(1)}(t)\rangle&=&\frac{ \vert\theta_{1}(t)\vert^{2}+\vert\theta_{0}(t)\vert ^{2}\cos \theta(t)}{\theta^{2}(t)}|0;0\rangle+\frac{i\sin\theta(t)}{\theta(t)}\sum_{\ell=\pm}\theta_{l,0}(t)e^{i\omega_{l,0}t}\vert \ell;0\rangle\nonumber \\
&&+\frac{\cos\theta(t)-1}{\theta ^{2}(t)}\sum_{\ell=\pm }\sum_{s=\pm }\theta _{s,0}(t)\theta _{s,\ell,1}(t)e^{i\omega_{\ell,1}t}\vert \ell;1\rangle,
\end{eqnarray}
where we define $\theta_0(t)=\sqrt{|\theta_{+,0}(t)|^2+|\theta_{-,0}(t)|^2}$ with $\theta_{\pm,0}(t)=\tilde{\mu}_0\int_{t_0}^tdt'\mathcal{E}(t')\exp(-i\omega_{\pm,0}t')$, and $\theta(t)=\sqrt{|\theta_0(t)|^2+|\theta_1(t)|^2}$ with $\theta_1(t)=\sqrt{|\theta_{+,1}(t)|^2+|\theta_{-,1}(t)|^2}$, by setting $\theta_{\ell',1}(t)=\sqrt{|\theta_{+,\ell',1}(t)|^2+|\theta_{-,\ell',1}(t)|^2}$ with $\theta_{s,\pm,1}=\tilde{\mu}_{s}\int_{t_0}^tdt'\mathcal{E}(t')\exp[i(\omega_{s,0}-\omega_{\pm,1})t']$.\\ \indent
\begin{figure}[h]
\resizebox{0.45\textwidth}{!}{%
\includegraphics{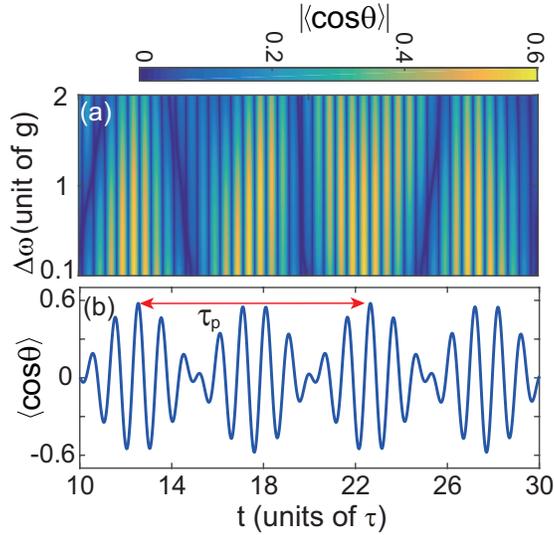} }\caption{The numerically calculated post-pulse orientation of the  polariton driven by an optimized laser pulse. (a) The time-dependent degree of orientation $|\langle\cos\theta\rangle(t)|$  versus the bandwidth $\Delta\omega$ of the laser pulses, (b) $\langle\cos\theta\rangle(t)$ at a narrow bandwidth of $\Delta\omega=0.1g$. The red arrows indicate the revival period $\tau_p$ of the polariton.}
\label{fig4}
\end{figure}
Based on Eq. (\ref{WFp1}), we conclude that the transitions to $|\pm,1\rangle$  can be perfectly blocked as long as $\mathcal{E}(t)$ satisfies the condition $|\theta_{\pm,1}(t_f)|=0$, reducing Eq. (\ref{WFp1}) into a $V$-type-three-level solution \cite{Shchedrin2015}.  It provides an alternative strategy to achieve the theoretical orientation maximum $\sqrt{3}/3$ by designing $\mathcal{E}(t)$ to  satisfy the following  conditions \cite{SM}
\begin{subequations} \label{apc}
\begin{align}
 &|\theta_{\pm,0}(t_f)|=\frac{\sqrt{2}\pi}{8},  \label{apca}\\
 &\omega_{-,0}\arg[\theta_{+,0}(t_f)]-\omega_{+,0}\arg[\theta_{-,0}(t_f)]=\pm g\pi.\label{apcb}
\end{align}
\end{subequations}
The amplitude condition by  Eq. (\ref{apca}) ensures optimal population distributions with $0.5$ in $|0;0\rangle$ and 0.25 in $|\pm,0\rangle$. The phase condition by Eq. (\ref{apcb}) enables that the faster coherence  between $|0;0\rangle$ and $|+;0\rangle$  constructively interferes with the slower one  between $|0;0\rangle$ and $|-;0\rangle$. \\ \indent%Thus,  the single pulse used in Fig. \ref{fig2} with a constant phase does not satisfy the two conditions simultaneously.
\begin{figure}[h]
\resizebox{0.55\textwidth}{!}{%
\includegraphics{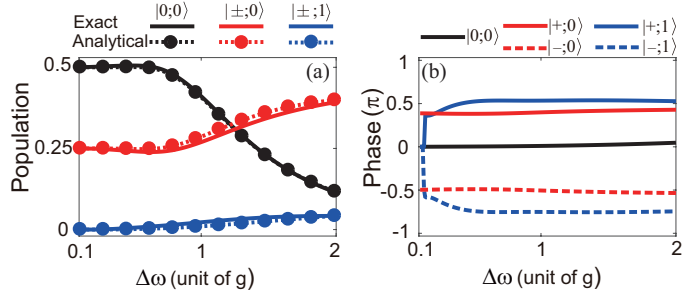} }\caption{The final populations (a) and phases (b) in  states $|0;0\rangle$, $|\pm;0\rangle$ and $|\pm;1\rangle$ versus the bandwidth $\Delta\omega$ of two laser pulses. The exactly solved numerical results (solid line) are compared with the analytical ones (dashed lines) in (a).}
\label{fig5}
\end{figure}
We perform numerical simulations  to determine the orientation maximum of the polariton. A composite laser field $\mathcal{E}(t)=\mathcal{E}_0f(t)\sum_{\ell=\pm}\cos(\omega_{\ell,0}t+\phi_{\ell,0})$ is applied in Eq. (\ref{Hp}) with choices of  $\mathcal{E}_0=\sqrt{2/\pi}A_1/(\tilde{\mu}_0\tau_0)$, $A_1=\sqrt{2}\pi/8$, $\phi_{-,0}=0$ and $\phi_{+,0}=\pi/9$, which satisfies  both the amplitude and phase conditions in Eq. (\ref{apc}).  Figure \ref{fig4} shows the dependence of $\langle\cos\theta\rangle(t)$ on the pulse bandwidth  $\Delta\omega$. The degree of orientation in the narrow-bandwidth regime can reach the maximum, i.e.,  0.5774, and decreases in the broad-bandwidth regime. As shown in Figs. \ref{fig4}(a,b), the orientation dynamics exhibits much more complex revival behaviours than the bare two-state molecule.  The  revival period of the polariton is significantly prolonged  to $\tau_p=10\tau$, in agreement with the theoretical value \cite{SM}. \\ \indent
Figure \ref{fig5} shows the final populations and phases in states $|0;0\rangle$, $|\pm;0\rangle$ and $|\pm;1\rangle$ versus the bandwidth $\Delta\omega$ of the composite laser fields.  Exact numerical results using the wave function in Eq. (\ref{WFp}) are compared with the analytical theoretical ones in Eq. (\ref{WFp1}). In the narrow bandwidth regime, the final populations as expected are localized in the lowest three states $|0;0\rangle$ and  $|\pm;0\rangle$ with optimal distributions, where the photon blockade effects take place without populating $|\pm;1\rangle$. As the bandwidth increases, the population transfers to  $|\pm;1\rangle$ in Fig. \ref{fig5}(a) become visible, and the phase values of the lowest three states  in Fig. \ref{fig5}(b) deviate slightly from the theoretical values that keep unchanged in the narrow bandwidth regime.  This implies that the condition $|\theta_{\pm,1}(t_f)|=0$ breaks down in the broad bandwidth regime, resulting in the dependence of orientation on the bandwidth in Fig. \ref{fig4}. The slight differences between the exactly and analytically calculated results in Fig. \ref{fig5}(a) may come from the transitions via higher-order Magnus terms, ignored in the analytical solution in Eq. (\ref{WFp1}) \cite{shu2020,shupra2021}.\\ \indent
To conclude, we contributed theoretical proof to complete quantum coherent control of rotational dynamics at a single molecular-polariton level. As a proof of principle, we performed simulations by considering a single-mode cavity in resonance with the lowest two rotational states of a single molecule driven by a  terahertz pulse with a narrow bandwidth. Our results illustrated that the presence of the cavity reduces the theoretical orientation maximum of the polariton and modifies its revival period. We derived an analytical solution of the pulse-driven quantum Jaynes–Cummings model by describing the polariton in terms of hybrid entangled states and utilized it to design the molecule-driving fields. We demonstrated that the theoretical orientation maximum of the polariton could be revived to the same value as the bare molecule by controlling the amplitude and phase parameters of a composite terahertz field.\\ \indent
To the best of our knowledge, this is the first exploration of the post-pulse orientation of molecules in a cavity, providing a direct signature of the RSC. This work offers a strategy for complete quantum coherent control of strongly-coupled molecule-cavity systems, a long-standing goal in chemistry and physics \cite{brumer1986,Zhang2017,brumer2021}. A natural extension of the present work would be to explore novel quantum optical phenomena of the molecular polariton in the ultrastrong, even deep-strong coupling regimes \cite{nrp2019,RMP2}. \textcolor{blue}{It would also be interesting to extend the present method to  a multi-molecule Jaynes-Cumming model \cite{nrp2019,RMP2,SekePRA,SekeOpt}, that can significantly reduce the requirement of the coupling strength of the cavity to each molecule. The experimental implementation of coherent control of molecular rotation in cavities requires the state of the art of micro-cavity techniques, for which molecule-cavity strong and ultrastrong coupling in the terahertz regime has been observed recently in different experiments \cite{THZE1,THZE2}. }
\begin{acknowledgements}
 The authors are grateful to Prof. Stephen Hughes of Queen's University for his  constructive comments and suggestions. This work was supported by the National Natural Science Foundations of China (NSFC) under Grant Nos. 12274470 and 61973317 and the Natural Science Foundation of Hunan Province for Distinguished Young Scholars under Grant No. 2022JJ10070. D. D. is supported in part by the Australian Research Council’s
Discovery Projects Funding Scheme under Project DP190101566, and the U. S. Office of Naval Research Global under Grant N62909-19-1-2129. F. N. is supported in part by
Nippon Telegraph and Telephone Corporation (NTT) Research, the Japan Science and Technology Agency (JST) via the Quantum Leap Flagship Program (Q-LEAP) and the Moonshot R\&D under Grant Number JPMJMS2061, the Japan Society for the Promotion of Science (JSPS) via the Grants-in-Aid for Scientific Research (KAKENHI) under Grant No. JP20H00134, the Army Research Office (ARO) under Grant No. W911NF-18-1-0358, the Asian Office of Aerospace Research and Development (AOARD) under Grant No. FA2386-20-1-4069, and the Foundational Questions Institute Fund (FQXi) under Grant No. FQXi-IAF19-06. L.-B. F. acknowledges financial support in part from the Fundamental Research Funds for the Central Universities of Central South University under Grant No. 1053320211611.
\end{acknowledgements}

\bibliographystyle{}

\end{document}